\documentclass[12pt,onecolumn]{revtex4}
\usepackage{graphicx}
\usepackage{bm}
\usepackage{epsfig}

\begin{document}

\title{Resonance energy transfer from a fluorescent dye molecule to plasmon and electron-hole excitations
 of a metal nanoparticle}
\author{R. S. Swathi and K. L. Sebastian \\
Department of Inorganic and Physical Chemistry\\
Indian Institute of Science, Bangalore 560012, India}
\date{\today}

\begin{abstract}
We study the distance dependence of the rate of electronic
excitation energy transfer from a dye molecule to a metal
nanoparticle. Using the spherical jellium model, we evaluate the
rates corresponding to the excitation of $l$ = $1$, $2$, and $3$
modes of the nanoparticle. Our calculation takes into account both
the electron-hole pair and the plasmon excitations of the
nanoparticle. The rate follows conventional $R^{-6}$ dependence at
large distances while small deviations from this behavior are
observed at shorter distances. Within the framework of the jellium
model, it is not possible to attribute the experimentally observed
$d^{-4}$ dependence of the rate to energy transfer to plasmons or
e-h pair excitations.
\end{abstract}
\maketitle

\section{\protect\bigskip Introduction}

Fluorescence Resonance Energy Transfer (FRET) is an interesting
photo-physical process \cite{stryer} that involves transfer of excitation
energy from donor to acceptor in a non-radiative fashion. It has been
extensively used in biology as a spectroscopic ruler to study the
conformational dynamics of biopolymers in the $10$-$100$ $\mathring{A}$
range. The rate of non-radiative damping of dye molecules near quencher
molecules is found to vary as $R^{-6}$, where $R$ is the distance between
the donor and the acceptor. In FRET, energy is transferred from a donor
molecule to an acceptor molecule via the dipole-dipole interaction. The
matrix element for the interaction is given by $V_{DA}\propto \frac{|\bar{\mu}%
_{D}||\bar{\mu}_{A}|}{R^{3}}$ where $\bar{\mu}_{D}$ and
$\bar{\mu}_{A}$ are the transition dipole moments of donor and
acceptor. Therefore, the rate of energy transfer, which is
proportional to the square of the interaction matrix element varies
as $R^{-6}$. The F\"{o}rster expression for the rate is given by
$k_{nr}=k_{r}\left( \frac{R_{0}}{R}\right) ^{6}$ where $k_{r}$ is
the radiative rate and $R_{0}$ is the F\"{o}rster radius, which can
be expressed as an overlap integral between the donor emission and
acceptor absorption spectra as $R_{0}\propto \left( \kappa ^{2}\int \frac{%
d\omega }{\omega ^{4}}I_{D}\left( \omega \right) \alpha _{A}\left(
\omega \right) \right) ^{\frac{1}{6}}$. The orientation factor
$\kappa ^{2}$ takes into account the effect of relative orientation
of the donor and acceptor transition dipole moments and is given by
$\kappa ^{2}=\left( \sin \theta _{A}\sin \theta _{D}\cos \left(
\varphi _{A}-\varphi _{D}\right) -2\cos \theta _{A}\cos \theta
_{D}\right) ^{2}$. The value of $\kappa ^{2}$ is usually taken to be
$\frac{2}{3}$ assuming rapid orientational averaging of the donor
within the lifetime of its excited state \cite{dale}.

However, there are reports of deviations from the conventional FRET
behaviour in the literature. Two possible reasons for this are the
breakdown of point dipole approximation, especially for systems with
extended transition densities like polymers \cite{Rossky} and the
incomplete orientational averaging within the lifetime of the
excited state \cite {lipman}. Currently, there is a lot of interest
in using nanoparticles \cite {alivi} as quenching agents due to
their tunable optical properties \cite {whetten}. There have been
very interesting observations on energy transfer between the dye
fluorescein and a gold nanoparticle of diameter $1.4$ $nm$.  Using
double stranded DNA as the spacer, the dye molecule was kept at
different fixed distances from the particle and the rate of energy
transfer was experimentally determined \cite{Strouse,Jennings}. They
found a $d^{-4}$ dependence of the rate on the distance $d$ between
the dye and the surface of the nanoparticle and have referred to
this as nanoparticle surface energy transfer (NSET). Such a
dependence is of great interest as it would more than double the
range of distances that can be measured. Persson and Lang (PL)
\cite{Persson} had long ago studied the dissipation of vibrational
energy of an oscillating dipole held at a distance $d$ above the
surface of a semi-infinite metal. In this case, the energy is
transferred to electron-hole (e-h) pair excitations in the metal,
which form a continuum, having
all possible energies from zero to infinity. For an excitation energy $%
\omega $, their density of states is proportional to $\omega $. PL
found the distance dependence to be $\sim d^{-4}$ (see also
\cite{chance,Whitmore}). Following this, it is suggested
\cite{Strouse,Jennings} that the observed distance dependence in
NSET is due to the excitation of e-h pairs in the nanoparticle. In
support of this, the authors \cite{Strouse,Jennings} point out that
plasmonic absorptions have not been observed for a gold nanoparticle
of diameter $1.4$ $nm$. However, unlike a semi-infinite metal, a
nanoparticle of such a small size, does not have a continuum of e-h
excitations. The excitations are discrete (see the excitation
spectrum given in Fig. \ref{fig2} of this paper). In view of this,
it would be very interesting to study the distance dependence of
NSET theoretically.

There have been attempts to explain the $d^{-4}$ dependence for the
case of nanoparticle \cite{singh,sangeeta,bagchi}. In particular,
focus of these papers \cite{bagchi} is on energy transfer to the
plasmons of the nanoparticle and they find a predominantly $R^{-6}$
dependence, though deviations occur at shorter distances. These
calculations do not account for e-h pair excitations. To the best of
our knowledge, there are no reports of theoretical calculations
which take plasmons as well as e-h excitations of the nanoparticle
into account. Therefore, we have calculated the rate of transfer of
electronic excitation from an excited fluorescein molecule to a
$1.4$ $nm$ nanoparticle, taking both the e-h pair excitations as
well as the plasmon excitations of the nanoparticle into account. We
model the nanoparticle within the jellium model \cite{brack}, and
use time dependent local density approximation (TDLDA) to calculate
the excitations of the nanoparticle.  Strictly speaking, one should
use RPA with the Hartree Fock approach as the starting point,
because in any LDA based approach, the single particle energies and
the response calculated using them have no fundamental meaning.
However, there have been a number of calculations based on TDLDA for
the response of a metallic nanoparticle and other systems
\cite{annrev,broglia,euro,theo}. All these calculations have led to
results which agree well with experiments and therefore we use this
approach. Within this framework, we do not get the experimentally
observed $d^{-4}$ dependence. Therefore, in our opinion, this very
interesting experimental observation is, as it stands, unexplained.
Perhaps, it may be due to other factors, like transfer through the
DNA, or the asphericity of the nanoparticle, which are being
currently investigated. It is also to be noted that the distances
between the donor and the acceptor in the NSET experiment are much
greater than the dimensions of the donor and acceptor, making it
unlikely that the breakdown of point dipole approximation is
responsible for the $d^{-4}$ dependence.

\section{Model for the nanoparticle}

We use the spherical jellium model for the nanoparticle which
provides a model system for investigating the response of the
conduction electrons in small metal particles
\cite{ekardt1,Bertsch}. In this model, the positive ions of the
metal cluster are replaced by a uniform sphere of positive charge
and the density functional formalism within the linear response
approximation is used to calculate the response of the cluster to
the time dependent external potential. Two kinds of excitations are
possible for a metal cluster namely, single particle excitations and
plasmon excitations. In a single particle excitation, an electron is
excited from an occupied level to an unoccupied level and these are
the electron-hole pair excitations. Plasmon excitations are
collective oscillations involving many electrons, wherein the
electronic charge density oscillates as a whole against the positive
background.

\begin{figure}[tbp]
\centering \epsfig{file=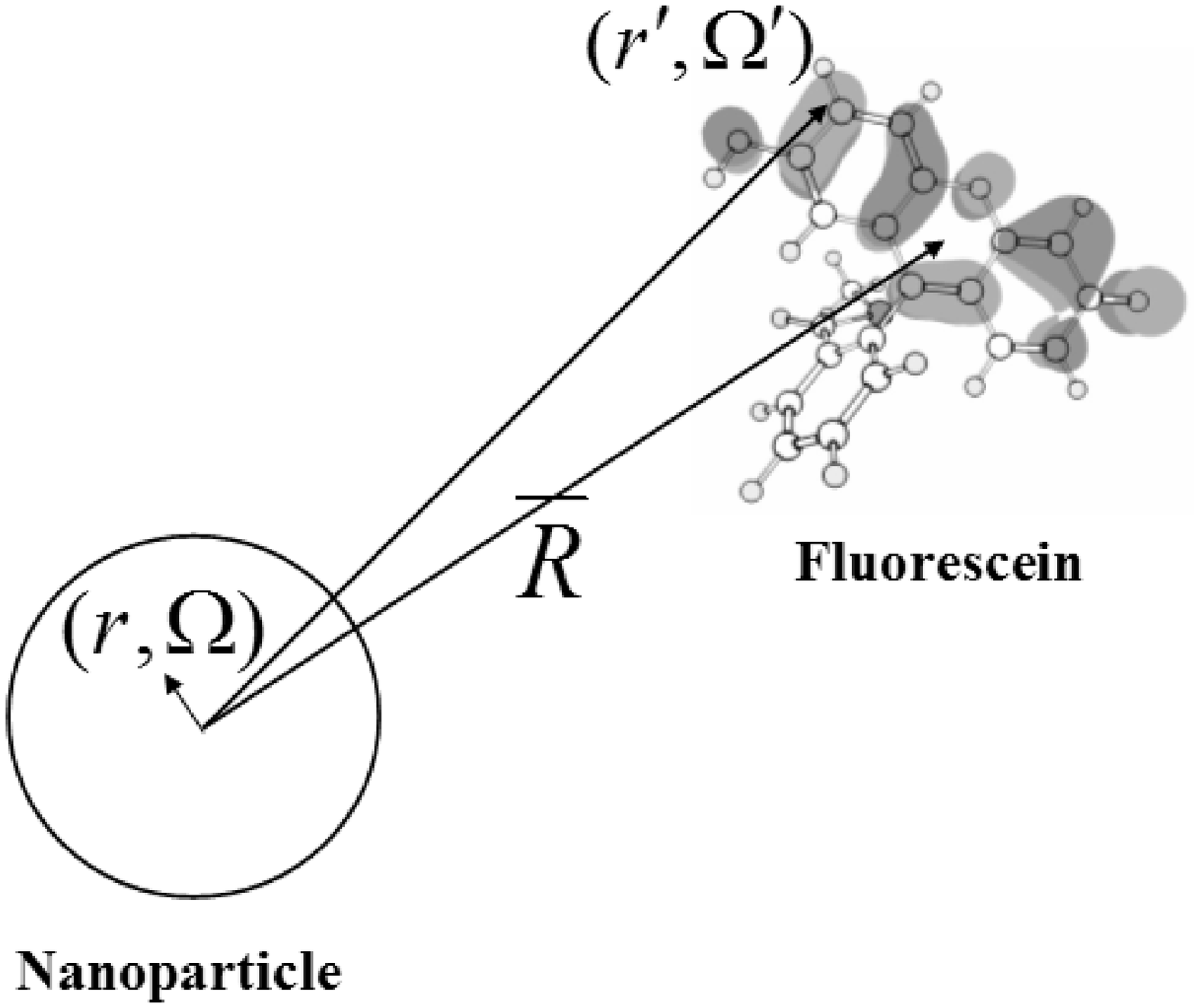,width=\linewidth}
\newline \caption{A schematic of the system consisting of the gold
nanoparticle and the dye molecule. } \label{fig1}
\end{figure}

\begin{figure}[tbp]
\centering \epsfig{file=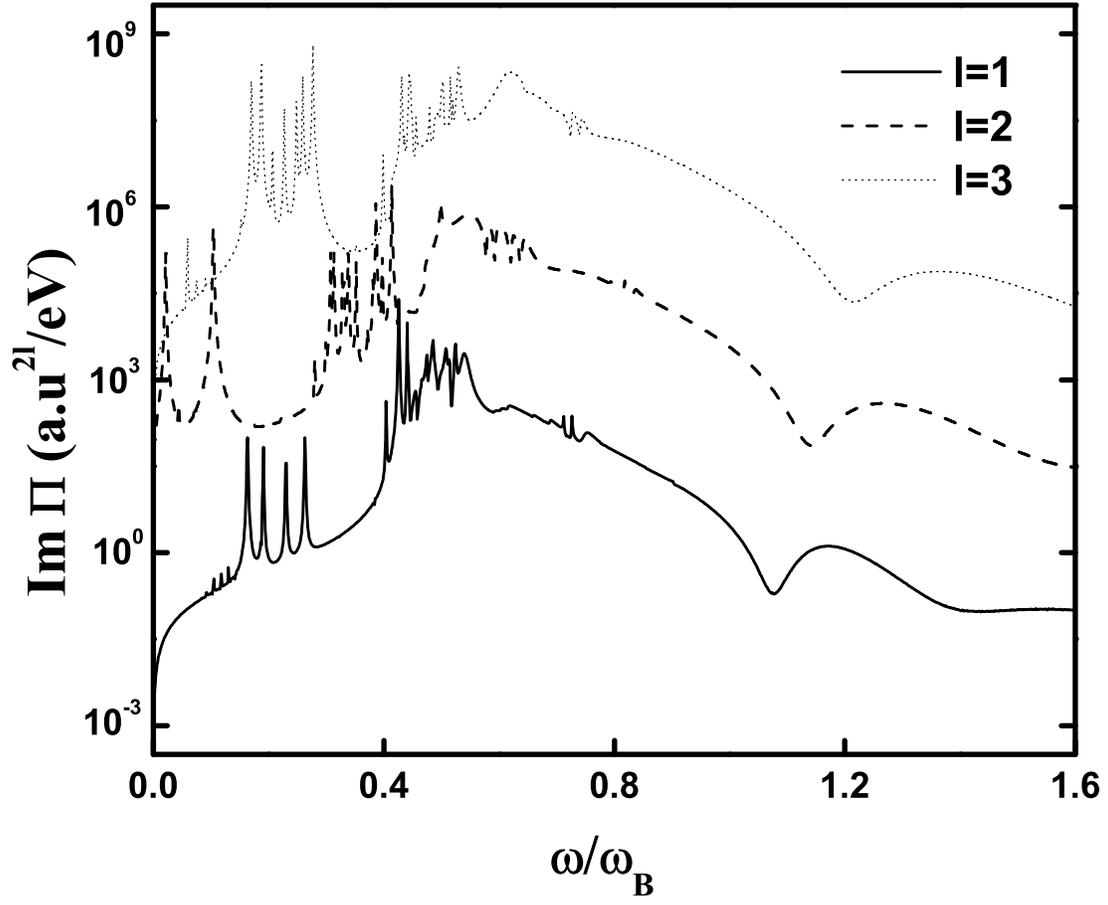,width=\linewidth}
\newline \caption{The imaginary part of the RPA polarization propagator for the
excitation of $l$ =$1$, $2$ and $3$ modes of the nanoparticle. The
frequency is given in units of the bulk plasmon frequency. }
\label{fig2}
\end{figure}

We denote the Hamiltonian of the nanoparticle as $H_{0}$. The
molecule is treated within a single particle model.  An electron
which was initially in the orbital $\phi _{g}$ is excited to an
orbital $\phi _{e}$.  De-excitation of the molecule, in which the
electron goes back to $\phi _{g}$ may be thought of as a time
dependent potential acting on the nanoparticle, given by
\begin{equation}
\Phi _{ext}(\bar{r},t)=e^{-i\omega t}\int \frac{\phi _{e}^{\ast }(\bar{r}%
^{^{\prime }})\phi _{g}(\bar{r}^{^{\prime }})}{|\bar{r}-\bar{r}^{^{\prime }}|%
}d\bar{r}^{^{\prime }}.
\end{equation}
In the above, $\omega $ is the frequency of the transition
corresponding to de-excitation. Thus the total Hamiltonian can be
written as:
\begin{equation}
H=H_{0}+\Phi _{ext}(\bar{r},t).
\end{equation}
One can carry out an expansion of the electrostatic potential at $\bar{r}$
due to a charge distribution at $\bar{r}^{^{\prime }}$ as \cite{Jackson}
\begin{equation}
\frac{1}{|\bar{r}-\bar{r}^{^{\prime }}|}=\sum_{l,m}\frac{4\pi }{2l+1}\frac{%
r_{<}^{l}}{r_{>}^{l+1}}Y_{l,m}^{\ast }(\Omega )Y_{l,m}(\Omega
^{^{\prime }}), \label{eqn3}
\end{equation}
where $r_{>}$ $(r_{<})$ is the larger (smaller) of $r$ and
$r^{^{\prime }}$ and $Y_{l,m}(\Omega )$ are the spherical harmonics.
From Fig. \ref{fig1}, it is obvious that the integration over
$\bar{r}^{^{\prime }}$ is to be performed over the molecule. As the
electron density of the molecule is fully outside the nanoparticle,
$r$ $<$ $r^{^{\prime }}$ and hence
\begin{equation}
\Phi _{ext}(\bar{r},t)=e^{-i\omega t}\sum_{l,m}M_{l,m}r^{l}Y_{l,m}^{\ast
}(\Omega )
\end{equation}
where
\begin{equation}
M_{l,m}=\frac{4\pi }{2l+1}\int \frac{\phi _{e}^{\ast }(\bar{r}^{^{\prime
}})\phi _{g}(\bar{r}^{^{\prime }})}{r^{^{\prime l+1}}}Y_{l,m}(\Omega
^{^{\prime }})d\bar{r}^{^{\prime }}.
\end{equation}
Thus the perturbation acting on the nanoparticle is a combination of
multipole fields of the form $r^{l}Y_{l,m}^{\ast }(\Omega )$ for
various values of $l$ and $m$. The case of such perturbations acting
on nanospheres of jellium is well studied in the literature
\cite{ekardt}.

\section{\protect\bigskip The rate of energy transfer}
The rate of transfer using the Fermi golden rule is:
\begin{equation}
k=\frac{2\pi }{\hbar }\sum_{E}|\left\langle E|\Phi _{ext}(\bar{r}%
,t)|G\right\rangle |^{2}\delta (E_{E}-E_{G}-\hbar \omega )
\end{equation}
where $|G\rangle $ is the initial (ground) and $\left|
E\right\rangle $ is the final (excited) state of the nanoparticle.
For a spherically symmetric closed-shell system \cite{beck}, within
the jellium model, the perturbation $r^{l}Y_{l,m}^{\ast }(\Omega )$
can only lead to the excitation of the $lm^{th}$ mode in the
nanoparticle. Thus, the excitations in the nanoparticle have the
same symmetry as the perturbation and the rate simplifies to
\begin{equation}
k=\frac{2\pi }{\hbar }\sum_{E,lm}M_{l,m}^{2}|\left\langle
E_{l,m}|r^{l}Y_{l,m}^{\ast }(\Omega )|G\right\rangle |^{2} \delta
(E_{E_{l,m}}-E_{G}-\hbar \omega ).
\end{equation}
$\left| E_{l,m}\right\rangle $ denotes excited states having quantum numbers
$l,m$.

If the nanoparticle is placed in an oscillatory external field $\Phi
_{ext}^{l}(r)e^{-i\omega t}= r^l e^{-i\omega t} $, an induced
electronic charge density is set up, which within the linear
response theory \cite{Bertsch,broglia} is given by
\begin{equation}
\delta \rho _{l}(r)=\int dr^{^{\prime }}\Pi _{l}(r,r^{^{\prime }},\omega
)\Phi _{ext}^{l}(r^{^{\prime }})
\end{equation}
where $\Pi _{l}(r,r^{^{\prime }},\omega )$ is the density-density
correlation function or the polarization propagator. Within the
random phase approximation (RPA), $\Pi _{l}(r,r^{^{\prime }},\omega
)$ obeys the following integral equation: \cite {Bertsch,broglia}
\begin{equation}
\Pi _{l}(r,r^{^{\prime }},\omega )=\Pi _{l}^{0}(r,r^{^{\prime
}},\omega )+  \int dr_{1}\int dr_{2}\Pi _{l}^{0}(r,r_{1},\omega
)\kappa _{l}(r_{1},r_{2})\Pi _{l}(r_{2},r^{^{\prime }},\omega )
\end{equation}
where $\Pi _{l}^{0}(r,r^{^{\prime }},\omega )$ is the independent
particle propagator and $\kappa _{l}(r_{1},r_{2})$ is the effective
two particle interaction \cite{Bertsch}. The independent particle
approximation to the response function $\Pi _{l}^{0}(r,r^{^{\prime
}},\omega )$, contains the e-h pair excitations only, while the
response under RPA, given by $\Pi _{l}(r,r^{^{\prime }},\omega )$
includes both the single particle and plasmonic response. The free
and RPA response to the external perturbation of the system is
calculated by integrating the external potential over the transition
density\cite{Bertsch,broglia}. Thus we have
\begin{equation}
\Pi _{l}^{0}(\Phi _{ext}^{l},\omega )=\int dr\Phi _{ext}^{l}(r)\delta \rho
_{l}^{0}(r)
\end{equation} and
\begin{equation}
\Pi _{l}^{RPA}(\Phi _{ext}^{l},\omega )=\int dr\Phi _{ext}^{l}(r)\delta \rho
_{l}^{RPA}(r)
\end{equation}
The polarization propagator is related to the strength function, $%
S_{l}(\omega )$ by
\begin{equation}
S_{l}(\omega )=\frac{1}{\pi }\mbox{Im}\Pi _{l}(\Phi _{ext}^{l},\omega ).
\end{equation}
The rate may be expressed in terms of the strength function $S_{l}(\omega )$
as
\begin{equation}
k=\frac{2\pi }{\hbar }\sum_{E,lm}M_{l,m}^{2}S_{l}(\omega )
\label{RateFinal}
\end{equation}
The evaluation of polarization propagator for the independent
particle as well as the RPA response allows one to take both the e-h
pair excitations as well as the plasmonic excitations into account
in the calculation.  We make use of the time dependent local density
approximation (TDLDA) version of the above equations, as this has
been found to lead to excellent results.  Details of the approach
may be found in \cite{broglia, Bertsch}.

\section{Calculations}

We have optimized the geometry of fluorescein, using the Gaussian03
program, within the DFT approximation (B3LYP-6-31G*). The HOMO
($\phi _{g}$) and LUMO ($\phi _{e}$) were taken from the calculation
for the optimised geometry. These orbitals were then used to
evaluate the matrix elements $M_{l,m}$ numerically for all $m$ with
$l$ up to $3$. A grid within a box of size $24\AA \times 24\AA
\times 24\AA $, within which the molecule was located was used for
this purpose. Note that $l$ = $1$ corresponds to the oscillation of
the electrons in the nanoparticle that has the shape of a p-orbital,
$l$ = $2$ that of a d-orbital and $l$ = $3$ that of an f-orbital. We
assumed an Au cluster of $90$ atoms which corresponds to a $1.4 $
$nm$ gold nanoparticle and performed the jellium model calculation
to evaluate $\Pi _{l}^{0}(\Phi _{ext}^{l},\omega )$ and $\Pi
_{l}^{RPA}(\Phi _{ext}^{l},\omega )$. The rate was calculated using
Eq. (\ref{RateFinal}), with static but random averaging over all the
orientations of the nanoparticle. Thus, we have evaluated the rate
of energy transfer from the fluorescein molecule to the $1.4$ $nm$
gold nanoparticle as a function of distance.

\section{Results and Discussion}

\begin{figure}[tbp]
\centering \epsfig{file=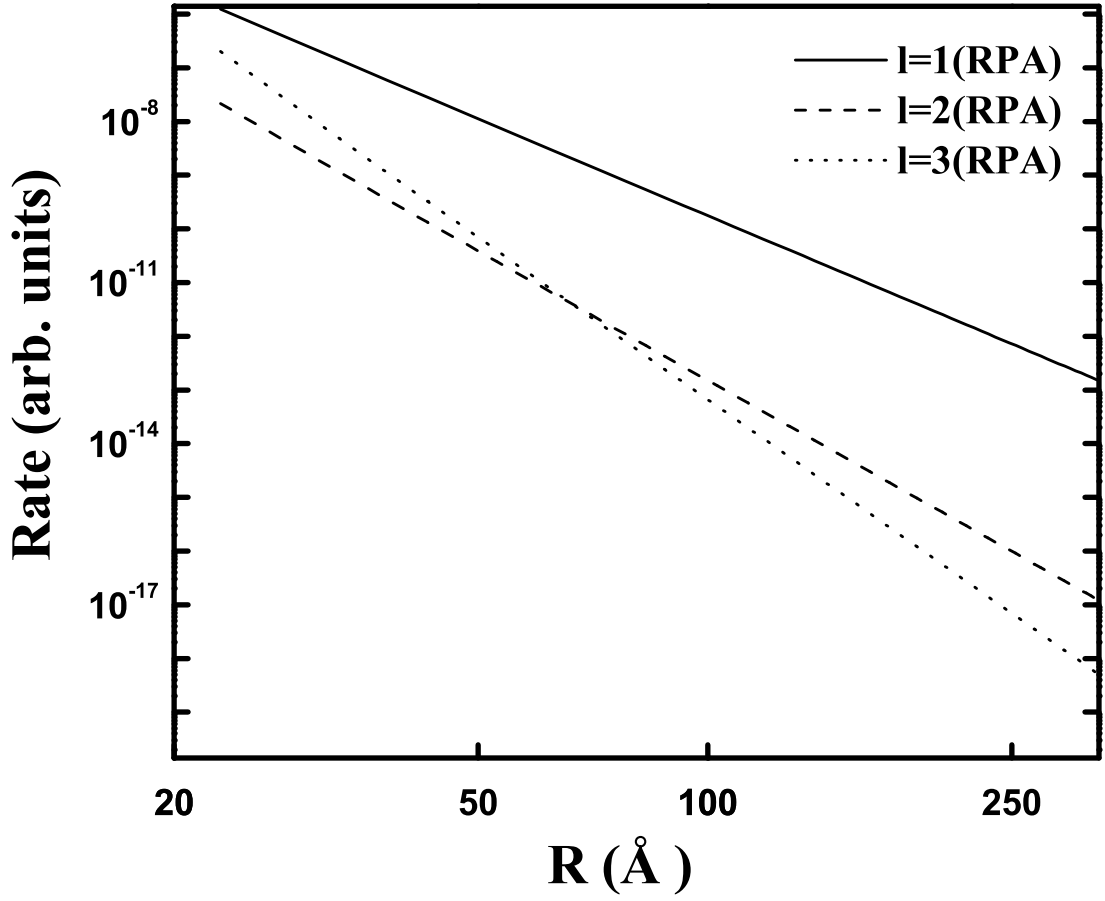,width=\linewidth}
\newline \caption{Distance dependence of the rate of transfer corresponding to
the excitation of $l$ =$1$, $2$ and $3$ modes of the nanoparticle
for the TDLDA response. The slopes of the log-log plots are
$-6.005$, $-8.019$ and $-10.039$ for $l$ =$1$, $2$ and $3$
respectively. } \label{fig3}
\end{figure}

\begin{figure}[tbp]
\centering \epsfig{file=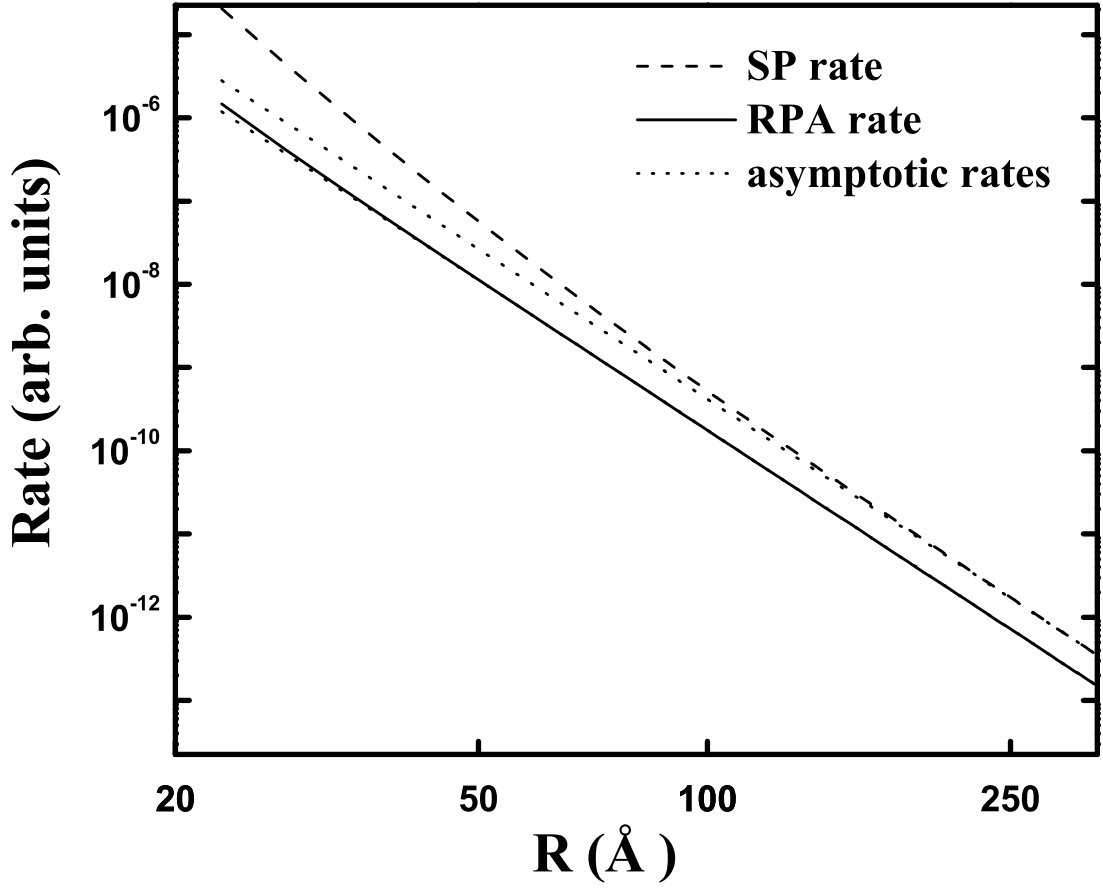,width=\linewidth}
\newline \caption{Distance dependence of the calculated total rate of transfer
for the single particle and RPA responses.  In both the cases, the
rate calculated using asymptotic expression $c/R^6$ (referred to as
asymptotic rates) are shown as dotted lines. The actual rate is
greater than this due to the contributions from $l=2$ and $l=3$
modes. } \label{fig4}
\end{figure}

The rate of energy transfer depends on the values of the two terms $M_{l,m}$
and $S_{l}(\omega )$. The first one represents the perturbing potential
acting on the nanoparticle due to the dye molecule and the second one is the
response of the nanoparticle to the perturbation. We have evaluated $%
S_{l}(\omega )$ for $l=1$, $2$ and $3$ modes of excitations of the
nanoparticle, for the TDLDA response using the JELLYRPA program of
Bertsch \cite{Bertsch}. For the calculation, we have taken $r_{s}=3$
atomic units (au). The broadening parameter $\Gamma $, which
determines the width of the single particle peaks was taken to be
$0.01$ $eV$. Plot of $S_{l}(\omega )$ calculated using the single
particle and RPA approaches is shown in Fig. \ref{fig2} for $l$ =
$1,2$ and $3$ modes. The TDLDA plot clearly shows peaks
corresponding to e-h pair excitations, collective surface plasmon
and bulk plasmon excitations. The numerous narrow spikes at lower
energies correspond to e-h pair excitations. The peak around $\omega
/\omega _{B}\simeq 0.5$ is the remnant of the surface plasmon and
the one around $\omega /\omega _{B}\simeq 1.2$ corresponds to the
bulk plasmon. Note that the frequency of a surface plasmon mode
\cite{Prodan} is related to
that of the bulk plasmon mode as $\omega _{S}=\omega _{B}\sqrt{\frac{l}{2l+1}%
}$. The surface plasmon mode is slightly red shifted while the bulk
plasmon mode is slightly blue shifted. The shifts are consistent
with previous jellium model calculations \cite{ekardt1,ekardt}. The
quantities of interest to us, in the calculation of the rate of
energy transfer to the nanoparticle are $S_{l}(\omega )$ for $l=1$,
$2$ and $3$, with $\omega $ having the value of emission energy of
the dye particle, which are given in Table 1 for $\omega =2.4eV$ .
\begin{center}
\begin{table}[tbp]
\begin{tabular}{|c|c|c|}
\cline{1-3}
$l$ & $S_{l}(\omega )$ (SP) & $S_{l}(\omega )$ (TDLDA) \\
& $(a.u.^{2l}/eV)$ & $(a.u.^{2l}/eV)$ \\ \hline
{$1$ } & {$32.69$ } & {$14.32$ } \\
{$2$ } & {$259500.$ } & {$322.4$ } \\
{$3$ } & {$5.714\times 10^6$ } & {$3.988\times 10^6$ } \\ \hline
\end{tabular}
\vspace{0.05in}
\caption{Numerical values of $S_{l}(\protect\omega )$ for $l$ =$1$, $2$ and $%
3$ at $\protect\omega =2.4eV$, the emission energy for fluorescein.
SP stands for single particle. } \label{table1}
\end{table}
\end{center}
After evaluating $M_{l,m}$ numerically for all $m$ with $l$
upto 3, we evaluated the rates of transfer corresponding to the $l$ = $1$, $%
2$ and $3$ modes of excitation of the nanoparticle, both for the
single particle as well as the TDLDA response. The results for TDLDA
are shown in Fig. \ref{fig3}. Note that the expansion in Eq.
(\ref{eqn3}) and hence the calculations are not valid if the dye
penetrates into the electron density of the nanoparticle. Moreover,
in such a case, overlap effects dominate and one has to think of the
Dexter mechanism and evaluate the exchange integral. The rates for
$l$ = $1$, $2$ and $3$ modes vary with distance as $R^{-6}$,
$R^{-8}$ and $R^{-10}$ as may be seen from the slopes in Fig.
\ref{fig3}. Then, we evaluated the total rate of transfer, again for
single particle as well as TDLDA response (see Fig. \ref{fig4}). It
is found from the figure that the single particle rate is actually
larger than the TDLDA result.
 This is because there is intensity borrowing by the collective modes (bulk and surface
plasmons) from the single particle excitations, resulting in a lower
value for $S_l(\omega)$ in the TDLDA calculations, as is clear from
table \ref{table1}. Asymptotically, at large distances, it is only
the $l$ = $1$ mode that is important and this leads to the
conventional $\sim R^{-6}$ limit for the total rate. But, at shorter
distances, $l$ = $2$ and $3$ modes gain importance leading to
deviations from $R^{-6}$ behavior at shorter distances. To make this
clearer, we fitted the long distance rate with $\frac{c}{R^{6}}$ and
used the result to calculate the short distance rate. The resultant
rate is shown in Fig. \ref{fig4}, along with the actual rate. The
actual rates are found to be \emph{slightly higher} than expected
from $\sim R^{-6}$ dependence. This is due to the fact that $l$ =
$2$ and $3$ modes become important at lower distances.\emph{ Note
that the deviations are not such as to give an $R^{-4}$ dependence
in the distance range that we study.} It is also of interest to note
that the $R$ dependence is governed by $\left| M_{l,m}\right| ^{2}$
and not by $S_{l}(\omega )$. Use of a different set of values of
$r_{s}$, or $\omega $ would not change the $R$ dependence.
Therefore, we conclude that excitation of plasmons or e-h pairs
cannot lead to the observed experimental data. In ref.
\cite{bagchi}, for a distance range from infinity to up to $4$ times
the radius of the nanoparticle (for our case, $28\;\AA$), the
behavior is found to be $R^{-6}$.  In this range, we too get an
$R^{-6}$ dependence, illustrating that inclusion of  e-h pair
excitations does not affect the result at all.  Also, it is to be
noted that in our calculations, our grid used for numerical
integration will penetrate into the nanoparticle for distances less
than about $28\; \AA$ so that our approach will break down for
closer distances.

\section{Conclusions}
We have adopted a spherical jellium model to evaluate both the
independent particle and collective response of a metallic
nanoparticle to an external time dependent perturbation and used
this to evaluate the rates of non-radiative energy transfer from the
excited state of a fluorescent dye to the particle. \ The rates due
to the excitation of $l=1,2$ and $3$ modes of the nanoparticle were
found to vary with distances as $R^{-6}$, \ $R^{-8}$ and $R^{-10}$
respectively.  The major contribution to the rate is from the
leading  $R^{-6}$ term.  The contributions from the other terms are
relatively small, except at distances close to $28 \AA$, and these
make the rate only larger and they are not in a direction as to make
the result behave like $d^{-4}$. Thus the experimentally observed
$d^{-4}$ dependence cannot be explained by considering the
excitation of plasmons or e-h pairs of the nanoparticle. \

\textbf{Acknowledgement}. R. S. Swathi acknowledges Council of Scientific
and Industrial Research (CSIR), India for financial support.

\newpage

\clearpage

\clearpage

\clearpage

\end{document}